\documentclass{appolb}

\usepackage[latin1]{inputenc}
\usepackage[dvips]{graphicx,epsfig,color}
\usepackage{psfrag,wrapfig,rotating}
\usepackage{amssymb,amsmath,array}

\def\slashchar#1{\setbox0=\hbox{$#1$}
   \dimen0=\wd0
   \setbox1=\hbox{/} \dimen1=\wd1
   \ifdim\dimen0>\dimen1
      \rlap{\hbox to \dimen0{\hfil/\hfil}}
      #1
   \else
      \rlap{\hbox to \dimen1{\hfil$#1$\hfil}}
      /
   \fi}

\newcommand{\be}{\begin{equation}}
\newcommand{\ee}{\end{equation}}
\newcommand{\eq}{\end{equation}}
\newcommand{\rb}{\underline{r}}
\newcommand{\kb}{\underline{k}}

\begin{document}
\title{A test of the BFKL resummation  at  ILC
\thanks{Presented at the ``School on QCD, low-x  physics, saturation and diffraction, Copanello (Calabria, Italy), July 1-14 2007.}%
}
\author{M.~Segond
\address{LPTHE \\
UPMC Univ Paris 06, Paris, France}
\and
L.~Szymanowski$^{1}$, S.~Wallon$^2$
\address{1- SINS \\
  Warsaw, Poland
\vspace{.1cm}\\
2- LPT\\ 
Universit\'e Paris-Sud, CNRS, Orsay, France}
}
\maketitle
\begin{abstract}
\vspace{-.5cm} We consider the exclusive production
of $\rho^0$ meson pairs in $\gamma^*\gamma^*$ scattering in the Regge limit of QCD as a probe of BFKL resummation effects and   we
show the feasibility of the measurement of this process at the ILC. 
\end{abstract}

\section{Collinear and $k_t$ factorizations of the process}


\psfrag{p1}[cc][cc]{$k_1$}
\psfrag{p2}[cc][cc]{$k_2$}
\psfrag{q1}[cc][cc]{$q_1$}
\psfrag{q2}[cc][cc]{$q_2$}
\psfrag{l1}[cc][cc]{}
\psfrag{l1p}[cc][cc]{}
\psfrag{l2}[cc][cc]{}
\psfrag{l2p}[cc][cc]{}
\psfrag{r}[cc][cc]{$r$}
\begin{wrapfigure}{r}{0.3\columnwidth}
\vspace{-.5cm}
\centerline{\includegraphics[width=0.3\columnwidth]{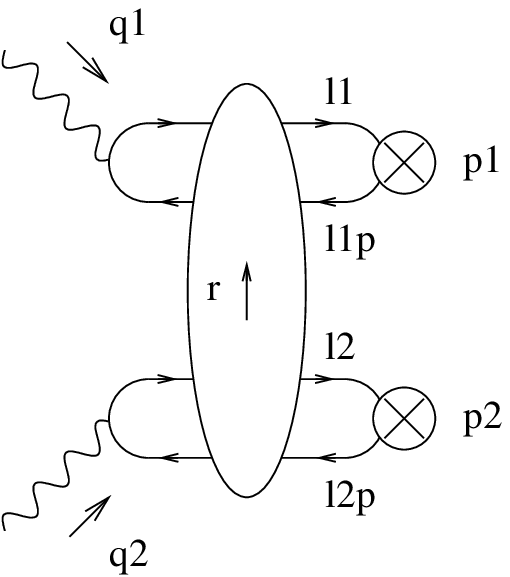}}
\vspace{-.50cm}
\caption{
}
\vspace{-.50cm}
\label{impact}
\end{wrapfigure}
In the high-energy (Regge) limit, when the cm energy $s_{\gamma^*\gamma^*}$ is much larger than all other scales of the process, large logarithms of $s_{\gamma^*\gamma^*}$ emerge and are resummed  by the BFKL equation \cite{bfkl}. It describes a $t-$channel hard pomeron exchange, made of a gluonic effective ladder and carrying the quantum numbers of the vaccuum. The highly virtual photons collision is a very clean process to probe the BFKL effects since it provides small transverse size objects ($q \bar{q}$ color dipoles) which overcome the theoretical problem arising from diffusion of the transverse momenta in the BFKL equation, at least in non asymptotical $s_{\gamma^*\gamma^*}$.  We can select  events with comparable photon virtualities to avoid the  partonic evolution of DGLAP \cite{dglap} type. Several studies \cite{bfklinc}  have been performed at the level of the $\gamma^*\gamma^*$ total cross-section and $J/\Psi$ meson pairs production in $\gamma\gamma$ collisions. Here we focus on the exclusive  process $\gamma^*_{L,T}(q_1) \gamma^*_{L,T}(q_2) \to \rho_L^0(k_1)  \rho_L^0(k_2)$  (see Fig.\ref{impact}) through $e^+e^-\to e^+e^-\rho_L^0\rho_L^0$ with double tagged outgoing leptons. The  $k_t$-factorization of the scattering amplitude, valid at high energy, has the form of a convolution in the transverse momentum  $\kb$ space between the  two  impact factors, corresponding  to the transition of
$\gamma^*_{L,T}(q_i)\to \rho^0_L(k_i)$ via the $t-$channel exchange of two reggeized gluons (with momenta $\kb$ and $\rb -\kb$). The  virtualities ($Q^2_i=-q^2_i $) of the photons supply the hard scale which justifies the use of perturbation theory to compute in the collinear factorization the hard part of each impact factor, convoluted with the soft part (encoding the hadronization into  the final states $\rho$ mesons) which is given by  the corresponding leading twist distribution amplitude (DA)\cite{BLphysrev24}. 
\vspace{-.110cm}

\section{Non-forward Born order cross-section }

\begin{wrapfigure}{r}{0.45\columnwidth}
\vspace{-.1cm}
\centerline{\includegraphics[width=0.45\columnwidth]{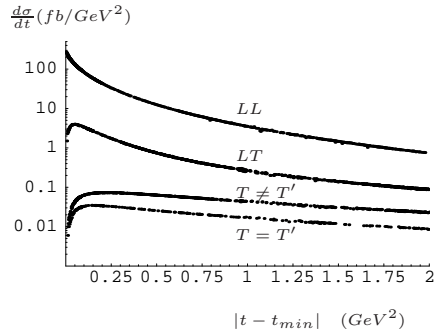}}
\begin{picture}(10,20)
\put(85,10){{\tiny $|t-t_{min}| $} \ {\tiny $ (GeV^2)$}}
\put(0,125){{\tiny $\frac{d\sigma}{dt} (fb/GeV^2)$}}
\put(86,90){\tiny $LL$}
\put(86,72){\tiny$LT$}
\put(86,58){\tiny$T\neq T'$}
\put(86,42){\tiny$T= T'$}
\end{picture}
\vspace{-.2cm}
\caption{{\footnotesize$e^+e^-\to e^+e^-\rho_L^0\rho_L^0$ cross-sections.}}
\vspace{-.2cm}
\label{FigLogcurves}
\end{wrapfigure}
We  display in Fig.\ref{FigLogcurves} the non-forward Born order cross-sections  as a function of the momentum transfer $t$ for the different $\gamma^{*}$ polarizations,  having  performed analytically the integrations over  $\kb$ (using conformal transformations to reduce the number of massless propagators) and numerically the integration over the accessible phase space \cite{born}. We then obtained the corresponding  cross-section of the process $e^+e^- \to e^+e^- \rho_L^0  \;\rho_L^0$ in the planned  experimental conditions 
of the International Linear Collider (ILC). We focus on the LDC detector project and we use the potential of the very forward region accessible through the electromagnetic calorimeter  BeamCal. Following the requirements of   Regge kinematics, we fix the value of $s_{\gamma^*\gamma^*}$  on  the gluon exchange dominance over the quark exchange contribution calculated in \cite{gdatda}. With the foreseen energy of the collider   $ \sqrt{s}=500$ GeV and nominal integrated luminosity of $125  \, {\rm fb}^{-1}$,  this will yield around $4 \cdot 10^3$ events per year, depending on the  theoretical asumptions we have made (scale dependence of $\alpha_s$, value of the parameter that controls the Regge kinematics and expansion of DAs).

\section{Forward differential cross-section with BFKL evolution}


The results obtained at Born approximation can be considered as the starting point for evaluation of the cross-section for $\rho^0$ mesons pairs production with complete BFKL evolution taken into account. We first evaluate  BFKL evolution in the leading logarithms approximation (LL) which 
dramatically enhances (by several orders of magnitude) the cross-section (and also the theoretical uncertainties coming mainly from the definition of the rapidity and the scale dependence of $\alpha_s$) when increasing $\sqrt{s}$,  because of the  large value of the LL BFKL Pomeron intercept.

\begin{wrapfigure}{r}{0.5\columnwidth}
\centerline{\includegraphics[width=0.5\columnwidth]{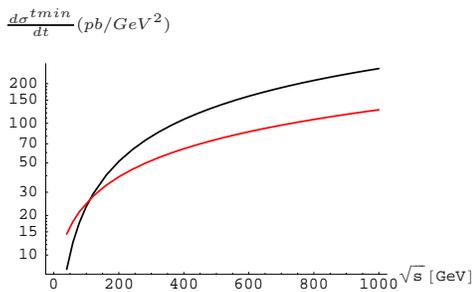}}
\begin{picture}(10,10)
\put(0,120){{\tiny $\frac{d\sigma^{tmin}}{dt} (pb/GeV^2)$}}
\end{picture}
\vspace{-.6cm}
\caption{{\footnotesize Cross-sections at $t=t_{min}$ for $\gamma^*\gamma^* \to  \rho_L^0  \;\rho_L^0$ with full NLL BFKL evolution (black) \cite{papa} and (this work) collinear improved BFKL evolution (red) for $Q_1=Q_2=2{\rm GeV}$ and three quark flavors.}}
\vspace{.4cm}
\label{FigNLL-Papatmin}
\end{wrapfigure}
\begin{wrapfigure}{r}{0.45\columnwidth}
\vspace{-6.95cm}
\centerline{\includegraphics[width=0.45\columnwidth]{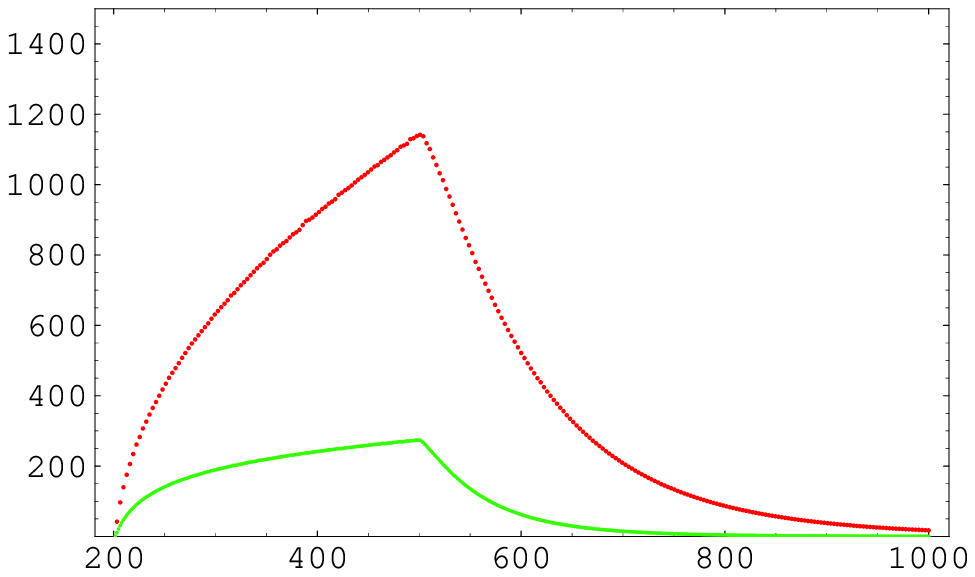}}
\begin{picture}(10,9.8)
\put(130,7){{\tiny $\sqrt{{\rm s}} $ \   $ [{\rm GeV}]$}}
\put(0,120){{\tiny$\frac{d\sigma^{tmin}}{dt} (fb/GeV^2)$}}
\end{picture}
\vspace{-0.4cm}
\caption{ {\footnotesize Cross-sections at $t=t_{min}$ for $e^+e^- \to e^+e^- \rho_L^0  \;\rho_L^0$ with collinear improved BFKL evolution   (red curve) and at Born order (green curve).}}
\vspace{-0.5cm}
\label{FigNLOtmin}
\end{wrapfigure}

The next-to-leading logarithms (NLL) BFKL evolution will  widely reduce both this enhancement and  uncertainties, which is essential to make precise predictions. The full NLL cross-section \cite{papa}, with both  impact factors and BFKL kernel computed in the NLL accuracy,  can even be lower at moderate values of $s_{\gamma^*\gamma^*}$ than its  Born order approximation. 
 We use the renormalization group improved BFKL kernel \cite{NLLpaper} (convoluted with LL impact factors) to estimate the NLL differential cross-section of  $\gamma^* \gamma^* \to \rho_L^0 \rho_L^0$, which gives a good agreement with the full NLL evolution obtained in \cite{papa}, as we can see in  Fig.\ref{FigNLL-Papatmin}. In the approach of Ref.\cite{epsw}, 
we must find the solutions (the NLL Pomeron intercept  and the anomalous dimension) of a set of two coupled equations (coming from the saddle point approximation and the residue of the integral over $\omega$, the Mellin moment of $s_{\gamma^*\gamma^*}$). Although this approach uses a fixed strong coupling, we reconstruct in $\omega_s$ and $\gamma_s$ a scale dependence by fitting with polynomials of  $Q_i$  a large range of solutions obtained for various values of $\alpha_s(\sqrt{Q_1 Q_2})$. 
Our results are now much less sensitive to the various theoretical asumptions  than the ones obtained at LL accuracy. Having integrated over the accessible phase space of this reaction at ILC, we compare in  Fig.\ref{FigNLOtmin} the curves  at Born order (green) with the (red) one obtained after collinear improved BFKL resummation. The experimental cut imposed by the resolution of the electromagnetic calorimeter BeamCal is responsible for the fall of the cross-sections with $\sqrt{s}$ increasing from $500$ GeV. This NLL evolution gives an enhancement of the Born approximation by a factor $4.5$, which allows us to definitively conclude of the measurability of the BFKL evolution for this process at ILC. We finally mention that increasing the collider energy from $500$ GeV to $1$ Tev will probably lead to a transition between the linear and the saturated regime ($Q_{sat} \sim 1.4$ GeV for $\sqrt{s}=1$ TeV).




\section{Acknowledgments}
 \vspace{-0.2cm}
We are very grateful to the organizers of the school. We thank D.~Y.~Iva-nov and A.~Papa for providing their curve, and R.~Enberg and B.~Pire for discussions and comments. We thank the ANR-06-JCJC-0084-02 for support. L.Sz. thanks the support of Polish Grant 1 P03B 028 28. 

\begin{footnotesize}

\end{footnotesize}

\end{document}